\documentclass[aps,prd,
superscriptaddress,groupedaddress,nofootinbib]{revtex4-2}

\usepackage{amsmath,amssymb}
\usepackage{epsfig,graphicx}
\usepackage{amssymb}
\usepackage{epsfig,color}
\usepackage{pstricks,graphicx,epsfig,color,amssymb,amsmath,amscd}
\usepackage{soul}

\usepackage{subfigure}
\graphicspath{
	{./Pics/}
}

\usepackage{braket}
\usepackage{newunicodechar}
\newunicodechar{⊙}{}

\usepackage{makeidx}
\newcommand{\bi}{\bibitem}
\newcommand{\be}{\begin{eqnarray}}
\newcommand{\ee}{\end{eqnarray}}
\newcommand{\rar}{\rightarrow}

\usepackage[]{caption}
\def\-g{\sqrt{-g}}
\renewcommand\rho{\varrho}

\usepackage{multirow}

\begin{document}

\title{\bf Asymmetric baryon capture by primordial black holes and 
baryon asymmetry of the universe}
\author{A.~D.~Dolgov}
\email{dolgov@fe.infn.it}
\affiliation{Novosibirsk State University \\
	Pirogova ul., 2, 630090 Novosibirsk, Russia  }

\author{N.~A.~Pozdnyakov }
\email{pozdniko@gmail.com}
\affiliation{Novosibirsk State University \\
	Pirogova ul., 2, 630090 Novosibirsk, Russia  }

\date{\today}

\begin{abstract}
We have refined our previously suggested scenario of generation of the cosmological baryon asymmetry 
through an asymmetric capture of baryons and antibaryons  by primordial block hole~\cite{DP-BH-BG}. It 
is found that in the limit of weak interactions of hypothetical heavy baryons with the primeval plasma the 
effect can be strongly enhanced and the observed magnitude of the  asymmetry can be obtained for a 
wide range of the model parameters.

 \end{abstract}


\maketitle


\section{Introduction \label{s-intro}}
 
The generally accepted mechanism of generation of the cosmological baryon asymmetry was
suggested by Sakharov~\cite{ADS-BG} in 1967.  He formulated the following three necessary conditions for the baryogenesis:\\
1. Violation of C and CP symmetries in particle physics.\\
2. Non-conservation of baryonic number $B$.\\
3. Deviation from thermal equilibrium in the early universe.\\
Out of these three Sakharov's principles the nonconservation of baryons remains yet unconfirmed  
by experiment, so scenarios of baryogenesis which can operate without assumption of baryon
nonconservation are certainly of interest.

The magnitude of the asymmetry expressed in terms of the present day number densities  
of baryons, antibaryons, and photons of the cosmic microwave background radiation
is equal to (see e.g. review~\cite{pdg}):
\be 
 \beta = \frac{ n_B - n_{ \bar B}} { n_\gamma}  \approx 6 \cdot 10^{-10}   
\label{beta}
\ee
where  $n_B$ and  $n_{\bar B}$  are respectively the number densities of baryons and antibaryons 
(note that today $n_{\bar B} \ll n_B)$,
$n_\gamma = 411(T_\gamma /2.73^o {\rm K})^3~{\rm  cm^{-3}}$, 
and $T_\gamma = 2.73^o {\rm K}$ is the 
present day temperature of the cosmic microwave background (CMB) radiation.
  
There are many different models of baryogenesis which with properly chosen parameters 
lead to the correct value of the asymmetry,
for reviews see refs.~\cite{AD-YaZ,ad1,ckn1,rs,ad2,RiTr:RPiB,jmc}. 

In our scenario excessive  antibaryons are hidden inside black holes, so one may
say that baryonic number is formally conserved. However, low mass black holes quickly
evaporate and disappear from our world without any trace. Thus baryon number is not conserved and
the global $U(1)$ symmetry associated with baryonic number is broken.
The breaking of all global symmetries, including that related to baryon number, by black holes is 
known for a long time. In  particular, this breaking can lead to proton instability~\cite{YaZ-decay}.
Though the life-time with respect to this decay is fantastically long, $\tau_p \sim 10^{45}$ years, 
one has to admit that B-conservation in particle physics does not save the proton life. In the
TeV gravity~\cite{ ADD} proton might decay almost instantly but spin and electric charge of proton
could suppress the virtual BH formation and very strongly increase the proton life-time~\cite{BDF}.

The idea that black hole evaporation might lead to
different numbers of particles and antiparticles in outer world belongs to 
Hawking~\cite{sh-bg}, while Zeldovich~\cite{yaz-bg} proposed a concrete mechanism
of its realization.
Possible generation of the cosmological baryon asymmetry in the process of { primoridal black hole (PBH)}
evaporation and an estimate of its magnitude
are described in the subsequent publications~\cite{carr-BG,turner-BG}.
The calculations of the baryon asymmetry of the universe in the frameworks of the Zeldovich 
scenario~\cite{yaz-bg} have been performed  in refs.~\cite{ad-bh1,ad-bh2}. 

These pioneering works were followed by now with a plethora of scenarios according to which 
cosmological baryon
asymmetry could be generated in the process of black hole {\bf evaporation}~\cite{BH-evap-BG-1,BH-evap-BG-2,BH-evap-BG-3,BH-evap-BG-4,BH-evap-BG-5,BH-evap-BG-6,BH-evap-BG-7,BH-evap-BG-8,BH-evap-BG-9,BH-evap-BG-10,BH-evap-BG-11}. 
In contrast to them the model presented in this paper is based on the novel
idea that baryogenesis could proceed in the process of asymmetric
{\bf capture} of baryons and antibaryons by primordial black holes. Here
the ideas suggested in our previous paper~\cite{DP-BH-BG} are further developed. In 
ref.~\cite{DP-BH-BG} we considered slow diffusion of heavy baryons and antibaryons to
primordial black holes because of their short mean free path in the primeval plasma. We have
shown that a non-zero value of the baryon asymmetry may be generated in this process, though
a strong fine-tuning was necessary to get a reasonable result.

Here we consider the scenario with a large mean free path of the accreting baryons and because
of that the model becomes much less constrained allowing for generation of the observed 
cosmological baryon asymmetry.

The main point of the present work is the difference of mobilities of heavy nonrelativistic baryons
in the primeval plasma predominantly consisting of relativistic matter. This difference can be induced
by the breaking of C and CP invariance, though CPT remains unbroken in the same way as the
partial decay width of particles and antiparticles may be different if a sufficient number of the 
decay channels are open, see e.g. \cite{AD-YaZ-rev}. A detailed discussion of the cross-section
difference is presented in our paper~\cite{DP-BH-BG}. Hence the accretion rate becomes 
different and antibaryons can accumulate inside PBHs leaving excess of 
baryons in our external space.

The paper is organized as follows. In the following Section we present and solve the equation of motion of nonrelativistic particles 
accreting  to a central gravitating body (BH) in expanding Friedmann universe. In 
Sec.~\ref{s-bar-gen} cosmological baryon asymmetry due to antibaryon capture by PBH  
is calculated and the values of the parameters are fixed to ensure the proper magnitude of the
asymmetry. Throughout this paper these particles are called $X$ and $\bar X$ for baryons and 
antibaryons respectively. In Sec.~\ref{s-asym}  we describe essential features of C and CP symmetries violation
necessary for the implementation of the considered scenario (for more detailed discussion see ref.~\cite{DP-BH-BG}).
In Sec.~\ref{s-concl} we conclude.

\section{Accretion to BH in expanding universe \label{eqn-of-mot}}

Equation of motion (geodesic equation)
for nonrelativistic particles, $X$, in the curved background created 
by a black hole in the Friedmann  space-time is derived e.g. in ref.~\cite{NLH} and has the form:
\be
\ddot r = \frac{\ddot a}{a} r  - \frac{r_g}{2 r^2} +\frac{L^2}{r^3},
\label{ddot-r}
\ee
where $r_g$ is the Schwarzschild radius, $r_g = 2M G_N \equiv  2 M/ m_{Pl}^2$, 
$M$ is the PBH mass, $G_N$ is the Newtonian gravitational constant,
$m_{Pl} =  1.22\cdot 10^{19}\, {\rm GeV} = 2.17\cdot 10^{-5}$ g, 
$L$ is the angular momentum the $X$-particle,
$a(t)$ is the cosmological scale factor, and the over-dots mean time derivatives. 
At the radiation dominated cosmological stage $\ddot a /a \approx - H^2 $ if
the PBH does not have an essential impact on the Friedmann expansion.
The angular momentum of the $X$-particle, $L$, is supposed to be zero since these  particles and
PBH are naturally at rest in the comoving frame.

We integrate this equation analytically assuming that $H$ slowly, adiabatically changes with time. 
As a result we find the following expression for the X-particle velocity, $\dot r$:
\be
\dot r^2 =  \frac{r_g}{r} - H^2 r^2 + v_1^2,
\label{dot-r-2}
\ee
where $v_1 = const$. {We fix this constant by the condition that the particle velocity
vanishes at the distance $r = r_{max}$ corresponding to the equilibrium between the Hubble repulsion and the 
gravitational attraction induced by the BH so $r_{max}$ is the  maximum radius of particle capture. As follows
from eq.~(\ref{dot-r-2}), $r_{max}$ is equal to
\be
r_{max}^3  = \frac{r_g}{H^2}.
\label{r-max}
\ee
So $v_1 = 0$ and eq. (\ref{dot-r-2}) is solved as
\be
\dot r = - \sqrt{ \frac{r_g}{r} - H^2 r^2 }
\label{dot-r-fin}
\ee
 
Equation (\ref{dot-r-fin}) can be further integrated resulting in:
 \be
 r (t) = r_{max} \left[\cos (3Ht/2) \right]^{2/3} .
 \label{r-of-t}
 \ee
 It is valid till $r(t)$ drops down to $r_g$. Since $r_g \ll r_{max}$, it can be reached at 
$3Ht/2$ close to $\pi/2$. It does not contradict our assumption of slow variation of $H$ which 
is true by an order of magnitude if $Ht \lesssim 1$.

 In the free fall approximation 
 there is no difference between the laws of motion for $X$ and $\bar X$. To take their difference into account we
 need to include into equation of motion (\ref{ddot-r}) a small friction term induced by the interaction of $X$ 
 and $\bar X$ particles with the plasma surrounding  a black hole. 
 With this correction the trajectory of $X$-particles, $R(t)$, would obey the equation:
   \be
\ddot R = \frac{\ddot a}{a} R  - \frac{r_g}{2 R^2} - \gamma \dot R ,
\label{ddot-R}
\ee
where $\gamma = \sigma_{el} v n_{rel}$,
$\sigma_{el}  \sim f^4 /m_X^2 $ is the cross-section of elastic $X$-particle scattering by the relativistic ones in cosmic plasma,
$f$ is the coupling constant, $v\approx 1$ is the relative velocity of $X$ {and a relativistic scatterer}, 
and $n_{rel} \approx 0.1 g_* T^3$ is the number density of all relativistic species in plasma, 
$g_* \approx 100$ is the number of relativistic species, and $T$ is the plasma temperature.
Due to C and CP violation $\gamma$ should be different for $X$ and $\bar X$, i.e.
$\Delta \gamma = \gamma_X - \gamma_{\bar X} \neq 0$.

Assuming that $\gamma/H$ is small,
we solve eq.~(\ref{ddot-R}) perturbatively expanding $R$ as $R= r + r_1$ where $r$ 
is the solution of the equation of motion (\ref{ddot-R}) in zeroth order in $\gamma$, so
$r(t) $ is given by expression~(\ref{r-of-t}). In the first order in $r_1$ eq.~(\ref{ddot-R}) is reduced to
the following linear inhomogeneous equation:
\be
\ddot r_1 = r_1 H^2 \tan^2 \left( \frac{3H t}{2} \right) - \gamma \dot r.
\label{ddot-r1}
\ee

It is convenient to introduce dimensionless time $\eta = H t$ and to rewrite eq.~(\ref{ddot-r1}) in the form:
\be
 r''_1 = r_1 \tan^2 \left( \frac{3 \eta}{2} \right) - \frac{\gamma}{H}  r',
\label{ddot-r1-dimensionless}
\ee
where prime means derivative over $\eta$.  By assumption $(\gamma /H) < 1$,

At short dimensionless time $r' \sim r_{max} \eta$ and $\tan^2 (3\eta/2) \approx 9\eta^2/4$,
so the first term at the right side of equation (\ref{ddot-r1-dimensionless}) appears to be smaller 
than the second one. Hence this equation can be solved as:
\be
 r'_1 = -(\gamma/H) \left( r_{max} - r (\eta) \right),
\label{dot-r-1}
\ee
where  $r_{max} $, given by eq. (\ref{r-max}), is supposed to be the initial value of $r(t)$.
An account of the first term in eq.~(\ref{ddot-r1-dimensionless}) does not change this result significantly. Thus 
$r_1(\eta)\approx ({\gamma}/{H}) r_{max} \eta $ or in other words 
\be
r_1 (t) \approx \gamma r_{max} t .
\label{r1-of-t}
\ee

The total number of $X$ or $\bar X$ particles captured by a BH during the Hubble time is approximately:
\be
N \approx (4\pi /3) \left(1 + 3 \gamma t \right) r_{max}^3 n_X,
\label{N}
\ee
where $n_X$ is the number density of $X$ particles. If the annihilation $X\bar X$ is weak (we check below
when it is indeed the case) and if
$X$-particles are efficiently produced by the inflaton decay at the end of inflation, then 
\be
n_X \approx n_{\bar X} \approx 0.1 g_s T^3 ,
\label{n-X-of-T}
\ee
where $g_s$ is the number of spin states of $X$-particles.

Since the $X\bar X$-annihilation is weak, the number density of $X$-particles remains unsuppressed
even at very small temperatures, $T < m_X$.

\section{Baryogenesis through capture of baryons by PBHs \label{s-bar-gen} }

We assume that the heavy particles $X$ and antiparticles $\bar X$ have non-zero baryon 
number $B_X\sim 1$. We also assume that  there exists an interaction between $X$, $\bar X$, and 
light particles which breaks C and CP symmetries but respects CPT. Some other
particles, heavier than $X$ and rather short-lived are also needed.
Their existence is necessary to create  difference between elastic cross-sections of X and 
$\bar X$ particles in the primeval plasma despite the CPT restrictions which demand
equality of the total cross-sections, see Sec.~\ref{s-asym}.

The accretion of X-particles to a PBH effectively started when these particles became
nonrelativistic. As we see in what follows, the smaller is the plasma temperature, the
larger is the baryon asymmetry, if the density of X-particles in comoving volume does not   
drop as $\exp (-m_X/T)$. In other words it happens if the $X \bar X$-annihilation froze at temperatures
of the order of $m_X$. However, such early freezing of massive species, if they are stable, would create
too high contribution into the cosmological density of dark matter.
The problem can be solved if $X$-particles are unstable but live sufficiently long to fulfill their task of creating the
baryon asymmetry of the universe. These and some other constraints are considered below
in this section.

Using equation (\ref{N}) we can estimate the difference between number of captured X
and $\bar X$ particles by a single PBH during time interval $t$:
\be
\Delta N \approx {4 \pi} r^3_{max} n_X t \Delta \gamma,
\label{N-per-BH}
\ee 
where $n_X$ is the number density of $X$-particles after they became nonrelativistic. We should keep in mind
that this time duration is bounded by the condition $t \lesssim H(T)$, where $T$ is the favorable temperature for
excessive $X$ over $\bar X$ capture by PBH, see below.

The difference between the friction coefficients in the case of maximally broken C and CP symmetries can be estimated as
\be
\Delta \gamma = \delta \sigma_{el} n_{rel} \approx f^6 n_{rel} / m_X^2,
\label{delta-gamma}
\ee
since the cross section of elastic scattering of $X$-particles on the relativistic particles is
$\sigma_{el} \approx f^4/m_X^2$ and  the difference between $X$ and $\bar X$ scattering is of the 
order of $f^2$
because the cross-section difference appears as a result of
radiative correction
proportional to $f^2$, exactly as there appears
the difference between partial decay widths
in the scenario of baryogenesis through massive particle decays.

This result
is true if the following conditions are fulfilled: the mean free path of X-particles in the primeval
plasma $l_{free}$ should be larger than the maximum capture radius $r_{max}$. The former can be estimated as:
\be
l_{free} = \frac{1}{\sigma_{el} n_{rel}} =  \frac{m_X^2}{0.1 g_* T^3 f^4 },
\label{l-free}
\ee
where $\sigma_{el} $ and $n_{rel}$ are defined  below eq.~(\ref{ddot-R}).

The condition $l_{free} > r_{max}$ can be rewritten as 
\be
\frac{m_X^2 H}{0.1 g_* f^4 T^3} > \left( r_g H \right) ^{1/3},
\label{l-free-to-r-max}
\ee
The Hubble parameter at the expansion stage dominated  by relativistic matter can be expressed through the temperature
applying the set of the following equations;
\be 
\rho = \frac{3 H^2 m_{Pl}^2}{8\pi} = \frac{3 m_{Pl}^2}{32\pi t^2} = \frac{\pi^2 g_*}{30}\, T^4 .
\label{rho-H-of-T}
\ee
Hence we obtain:
\be 
H = \left(\frac{8 \pi^3 g_*}{90} \right)^{1/2} \,\frac{T^2}{m_{Pl}} = 
16.6\, g_{100}^{1/2}\, \frac{T^2}{m_{Pl}},
\label{H-of-T}
\ee
where $g_{100} = g_*/100$.

Thus the bound (\ref{l-free-to-r-max}) can be rewritten as:
\be
M <\frac{0.14 m_X^6}{g_{100}^2 f^{12}\, T^5}
\label{bound-1}
\ee
Note that the limit does not depend upon the Planck mass.

If $m_X = 3\cdot 10^{13} $ GeV (close to the typical heating temperature after inflation), 
$T=m_X$, and $f = 0.1$,
the condition of the free fall is fulfilled for $M \lesssim 7$ g. For higher mass of the PBHs the free fall
condition is satisfied at smaller $T$, e.g. if $M= 10^6$ g, the efficient free fall capture took place at 
$T = m_X /10$. 

\bigskip

The rate of the annihilation is determined by the equation:
\be
\Gamma_{ann} \equiv \dot n_X /n_X = \sigma_{ann} v n_X = 0.1g_*\, g_s\, f_{ann}^4 T^3 / m_X^2,
\label{dot-nX}
\ee
where $f_{ann}$ is the coupling constant of the annihilation. Demanding that $\Gamma_{ann}$ is small in comparison with  
$H$, see eq.~(\ref{H-of-T}), we find that the annihilation would be inefficient at the temperatures
satisfying the condition:
\be
\frac{T}{m_X} < 2.5\cdot 10^{-6}  f_{ann}^{-4}\, \left(\frac{m_X}{3\cdot 10^{13} \rm{GeV}}\right).
\label{T-over-mX}
\ee
If the maximum value of $T/m_X$ may reach unity, then the  annihilation does not essentially diminish the density of $X$-particles below $n_X = 0.1 g_s T^3 $. In other words the density of X and $\bar X$
particles would be conserved in the comoving volume, i.e. $n_X = 0.1 g_s T^3 $ below $T= m_X$.
For $m_X = 3\cdot 10^{13}~\rm{GeV}$ this could be realized if $f_{ann}  \lesssim 4\cdot 10^{-2}$. Otherwise
the density of $X$-particles would be exponentially suppressed, $n_X \sim \exp (-m_X /T)$ at low temperatures.
To avoid an overclosing of the universe by $X$-particles we  assume that they are unstable, presumably
decaying before the Big Bang Nucleosynthesis.

\bigskip

Another important restriction on the efficiency of
the discussed here mechanism is that the "size" of X-particles i.e.
its Compton wave length $\lambda_X$ should be smaller than the gravitational radius of PBH. Otherwise 
the probability of the particle capture  would be suppressed by a power of the ratio $r_g/\lambda_c$:
\be
\lambda_X = 1/m_X < r_g = 2 M /m_{Pl}^2. 
\label{bound-2} 
\ee
It leads to a lower bound on the PBH mass:
\be
M > \frac{m_{Pl}^2}{2 m_X} = 4.4\, {\rm g} \left(\frac{3\cdot 10^{13} \,{\rm GeV}}{m_X} \right). 
\label{M-low-limit}
\ee

\bigskip

The baryon asymmetry gained by the PBH antibaryon capture can be diluted by the entropy 
release from  the PBH evaporation. As it follows from Ref.~\cite{ac-ad} this would be avoided if
\be
\epsilon M < 10^{-5}\, \rm{g} ,
\label{bound-3}
\ee
where $\epsilon$ is the fraction of the energy density of PBHs at the moment  of their formation:
\be
\frac{\rho_{PBH} (t_{form}) }{\rho_{rel} (t_{form}) } = \epsilon ,
\label{epsilon}
\ee
where $\rho_{rel} \approx 3 T  n_{rel} $ 
is  the energy density of the relativistic matter, and 
 \be
 t_{form} = {M}/{m_{Pl}^2} . 
 \label{t-form}
 \ee
 
 Using eqs.~(\ref{rho-H-of-T}) and (\ref{H-of-T}) we find that 
 the temperature of the relativistic matter at the formation moment is
 \be
 T_{form} \equiv    T(t_{form} ) = 0.17g_{100}^{-1/4} m_{Pl} \left(\frac{m_{Pl}}{M}\right)^{1/2}
 = 10^{14} \,{\rm GeV}  g_{100}^{-1/4} M_4^{-1/2},
 \label{T-form}
\ee
 where $M_4 = M /10^4$ g.  For successful baryogenesis PBHs should be created while X-particles are abundant in 
 the cosmological plasma. If $X \bar X$-annihilation continued till $T\ll m_X$, then the temperature of the PBH creation
 should be not much smaller than $m_X$. If, as we assume in the present work, the annihilation of $X\bar X$ in thermal
 plasma was never efficient, then PBHs should be produced prior to the decay of $X$ (and $\bar X$)-particles.
 
 In the course of the cosmological expansion and cooling down the energy fraction of PBH
 rises as $(T_{form}/T)$ until their evaporation, which happens at the time moment equal to the
 BH life-time $\tau_{BH}$~\cite{tau-BH}:
 \be
 t = \tau_{BH} \approx 30 M^3 /m_{Pl}^4 .
 \label{tau-BH}
 \ee
 The corresponding temperature is
 \be
 T(\tau_{BH} ) = 0.03 g_{100}^{-1/4}  m_{Pl}  \left(\frac{m_{Pl}}{M}\right)^{3/2} =
 3.7\cdot 10^4 \,{\rm GeV} M_4^{-3/2} .
 \label{T-of-tau}
 \ee
 The temperature of the relativistic plasma at the moment of PBH decay should
be smaller than $m_X$ to allow nonrelativistic X-particles to be captured by PBHs.
 
Since PBH are nonrelativistic, while the bulk of the matter is relativistic,
the fraction of the PBH energy density at temperature $T $ becomes larger than $\epsilon$
by the factor 
\be
\frac{T_{form}}{T} \approx g_{100}^{-1/4} M_4^{-1/2} \,\frac{10^{14}\,{\rm GeV}}{T}.
\label{T-form-to-T}
\ee

Now using Eq.~(\ref{N-per-BH}) and Eq.~(\ref{delta-gamma}) we find for the excessive baryon number create by a single primordial
black holes:
\be
\Delta N = 4\pi f^6 \,\frac{r_g n_{rel} n_X t }{H^2 m_X^2} = 
\frac{0.9 g_s f^6}{\pi^2} \,\frac{M T^2 t}{m_X^2}. 
\label{Delta-n2}
\ee

Hence the baryon asymmetry can be estimated as 
\be
\beta = \frac{ B_X n_{BH} \Delta N}{n_{rel}} = \frac{2.7}{\pi^2} B_X g_s f^6\,
\frac{\rho_{BH}}{\rho_{rel}}\frac{T^3 t}{m_X^2} =
 \frac{2.7}{\pi^2} B_X g_s f^6\,\epsilon \frac{T_{form} T^2 t}{m_X^2}
\label{beta-2}
\ee
where $\epsilon$ is the fraction of PBH energy density to that of the relativistic matter 
at the moment of their formation, see eq. (\ref{epsilon}),
$T_{form}$ is the temperature at PBH formation (\ref{T-form}),
and and $t \sim 1/H$, so we finally obtain:
\be
\beta \approx  0.016 B_X g_s f^6 \epsilon\, \frac{T_{form}}{T} \frac{m_{Pl} T}{m_X^2} 
\label{beta-fin}
\ee
Taking the maximum allowed values of $\epsilon$ from Eq~(\ref{bound-3}) 
$\epsilon = 10^{-5} g/ M$, $m_X = 10^{13}$ GeV, $f= 0.1$, $T= m_X/10$
$M=10^4 $ g and thus $T_{form} =10^{14}$ GeV we find that the baryon
asymmetry can easily reach the observed value and even overcome it. This choice of the
parameters satisfies the derived above restrictions.



{\section{Difference between mobilities of $X$ and $\bar X$ particles in the background 
plasma \label{s-asym}} }

In this section we revisit the main concepts from the corresponding section in our previous paper~\cite{DP-BH-BG}. 
First, we remind that we assume validity of  the first Sakharov condition of violation of  C and CP
symmetries, while the sacred CPT-invariance remains unbroken.

 As we have shown in ref.~\cite{DP-BH-BG}, in this case there should naturally appear difference between the 
 probabilities of the quasi-elastic scattering of $X$-particles over relativistic species in the cosmic plasma
 \be
\sum_{a,b} \Gamma ( X +a \rar X + b) \neq \sum_{\bar a, \bar b} \Gamma ( \bar X +\bar a \rar \bar X + \bar b),
\label{sigma-q-barq}
\ee
where summations are done over all light particle sets in the initial state $a$ and the final state $b$.

However, it should be taken into account that according to the CPT theorem the total probability 
$\Gamma [ {\bf p}_1, \lambda_1, a_1,$  ${\bf p}_2, \lambda_2, a_2, ...]$ of a 
process from an initial state, containing  a certain set of particles 
is equal to the total probability $\Gamma 
[ {\bf p}_1, -\lambda_1, \bar a_1, {\bf p}_2, -\lambda_2, \bar a_2, ... ]$ of the process containing antiparticles with opposite spin projection state, e.g.
with opposite helicities $\lambda$ see textbook~\cite{SW-QFT}, eq. (3.6.15), or review~\cite{AD-YaZ}:
\be
\Gamma \left[ {\bf p}_1, \lambda_1, a_1, {\bf p}_2, \lambda_2, a_2, ... \right] =
\Gamma \left[ {\bf p}_1, -\lambda_1, \bar a_1, {\bf p}_2, -\lambda_2, \bar a_2, ... \right] .
\label{Gamma-tot}
\ee

In particular, this condition leads to the mentioned above equality of the total decay widths of particles 
and antiparticles while allows for
a difference between the partial decay rates as well as for a difference among the partial modes of scattering 
processes, but the magnitude of the latter would be suppressed because the difference may appear only in 
higher order of perturbation theory.

Then to achieve the desired difference of $X$ and $\bar X$ particles scattering 
 we need to introduce a new interaction leading to
disappearance of $X$ and $\bar X$ particles through the process of the kind
\be
a + X \rar b + Y,
\label{AX}
\ee
where the heavy particle $Y$ may have zero baryonic number and
the light state $b$ should have the same baryonic number as 
$a + X$, if we want to avoid non-conservation of baryons. 

In complete analogy with the higher order corrections to the decay process, which create a difference 
between the partial decay widths of particles and antiparticles, we consider radiative corrections to elastic scattering 
of $X$ on relativistic particles, which lead to different values of the cross-sections. For more detail 
see our earlier paper~\cite{DP-BH-BG}. The corresponding Feynman diagrams are prsented 
in Figs.~\ref{fig:Xsc} and \ref{fig:Xresc}  an example of the process leading to 

\begin{figure}[ht]
	\centering \subfigure[]{
		\includegraphics[width=0.25\linewidth]{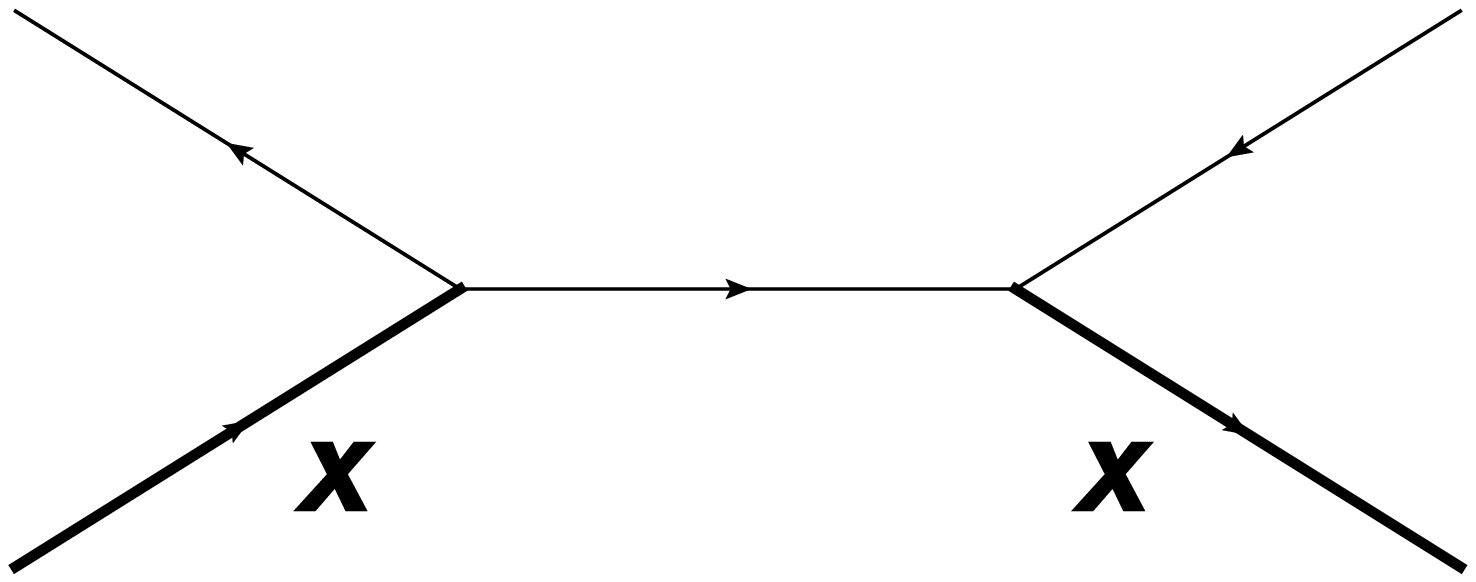}
		\label{fig:Xsc} }
	\hspace{4ex}
	\subfigure[]{
		\includegraphics[width=0.25\linewidth]{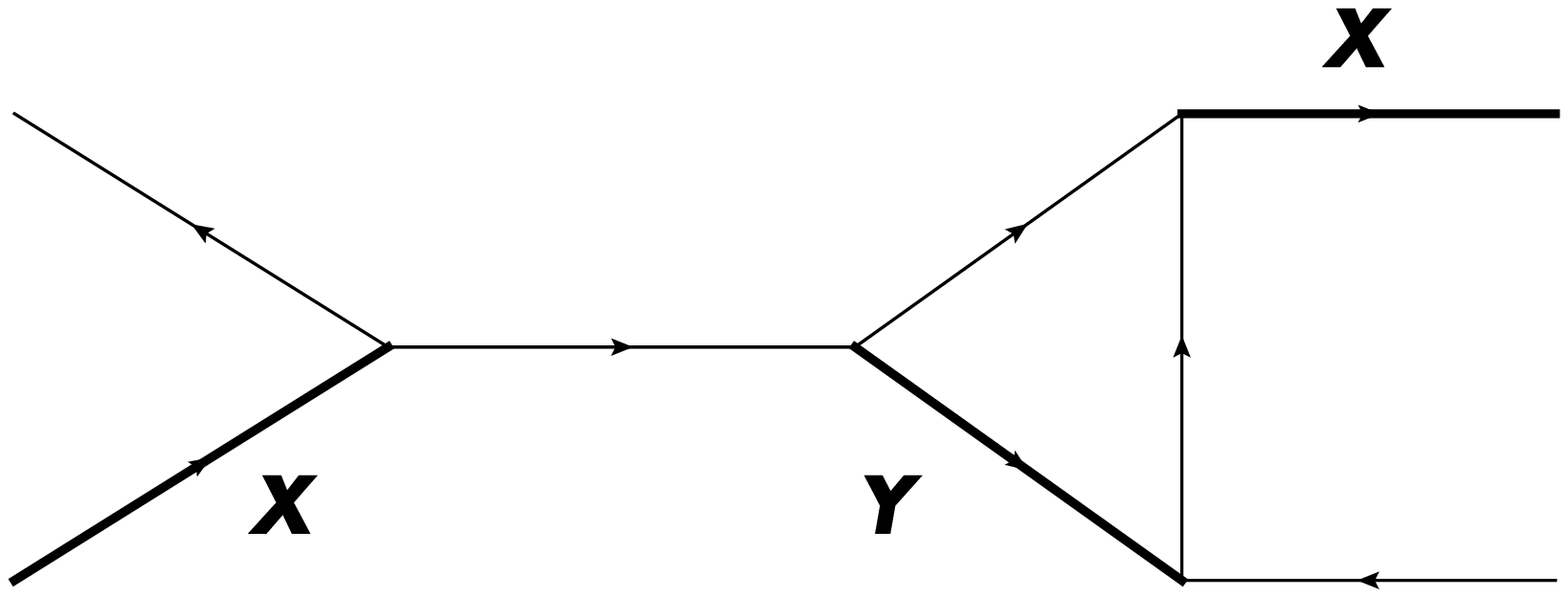}
		\label{fig:Xresc} }
	\caption{{Feynman diagrams describing $X$ (or $\bar X$)
			scattering off light quark. The first, 'a', diagram is
			the lowest order contribution and gives equal values for particles and antiparticles.  
			The 'b' diagram present an example of the one-loop correction with an exchange of $Y$-particle. 
			The one-loop scattering may be also present to the initial state and its contribution 
			multiplies the result by factor 2.
		}}
	\label{fig:2}	
\end{figure}

The equalities of the total probabilities means in particular that the total cross-section of $X$-scattering on particle {$"a"$}   
$\sigma_{tot} (X+a \rar All) $ is equal to the same of particles and antiparticles 
$\sigma_{tot} (\bar X+ \bar a \rar All) $. 
If the final state "All" contains one and only one $X$-particle, then the mobilities of $X$ and $\bar X$ in the cosmological
plasma would be the same and the discussed here mechanism of baryogenesis would not operate.
However, if the complete set of the final states includes a state or states where 
$X$ is missing (and analogously for the reactions with $\bar X$),
then the cross-sections of the processes $\sigma_{tot} (X+a \rar X+ All) $ and  
$\sigma_{tot} (\bar X+ \bar a \rar  \bar X +All) $
may be different, leading to the needed mobility differences of $X$ and $\bar X$.

The difference between probabilities of charge conjugated processes can be estimated as:
\be
\delta = \frac{\sigma_{X0}-\sigma_{\bar X 0}}{\sigma_{X0}+\sigma_{\bar X 0}} \approx \frac{|g_{x1}|^2Im(D)Im(g_{x1}g_{x2}^*g_{y1}g_{y2}^*)}{|g_{x1}|^4} \propto f^2 ,
\ee
where $\sigma_{X0}$ and $\sigma_{\bar X0}$ stand respectively
for the scattering cross-section of $X$  and $\bar X$ particles on relativistic particles in cosmic plasma.
The complex coefficient $D$ comes from the integration over 
the loop, and
$g_{xi}$ and $g_{yi}$ are partial decay constants of $X$ and $Y$ particles respectively.

The following supersymmetry  inspired model can serve as appropriate frameworks for the scenario.
Assume that $X$ is an analogue of the lightest supersymmetric particles (LSP)
which is stable due to an analogue of $R$-parity. Let assume that there exists
a heavier partner $H$ with zero baryonic number which would be unstable and decay through the 
channel $H \rar X+ 3 q$, where $q$ are light quarks with proper quantum numbers.
Accordingly the reaction $X + q \rar H + 2 \bar q$ becomes possible. It is exactly what we need to allow
for a difference between the cross-sections of the reaction $\sigma_{tot} (X+a \rar X+ All) $ and 
$\sigma_{tot} (\bar X+ \bar a \rar  \bar X +All) $, which can lead to a different mobilities of $X$ and $\bar X$
around black hole and to dominant capture of antibaryons over baryons creating cosmological baryon asymmetry.

Note that in $R^2$ gravity, Srarobinsky inflation~\cite{aas-infl}, the allowed mass of LSP-kind particle can
be close to $10^{13}$ GeV or even higher~\cite{EA-AD-RS-1,ADS-symmetry}. 

\section{Conclusion \label{s-concl} }

In this paper we continue investigation of baryogenesis through the asymmetric capture of baryons and antibaryons 
by primordial black holes. Unlike in our previous paper~\cite{DP-BH-BG}, where we used diffusion approximation 
$\gamma/H \gg 1$ in which particles many times scatter off the cosmic plasma before they are captured by PBH, 
in this study we investigate the opposite limit in which $\gamma/H \ll 1$ or the free fall limit. As it appears, there is a 
sufficiently wide parameters space to explain the observed value of baryon asymmetry of the universe.

A noticeable increase of the baryon asymmetry generated by the capture of the antibaryonic number by PBH
in the considered version of the scenario is achieved due to assumed negligible annihilation of $X$-particles
with decreasing temperatures, $T< m_X$, because at smaller $T$ the relative fraction of PBH with respect to the
total cosmological energy density goes up quite significantly.

The proposed here mechanism of baryogenesis does not demand two out of three Sakharov's principles. 
Namely it can proceed in thermal equilibrium and without assumption of non-conservation of baryonic number in 
particle reactions. It helps to avoid a possible problem which arises because non-conservation of baryons is not (yet)
observed in direct experiment

In a sense black holes break conservation of baryonic number, either hiding baryons in internal space making
them unobservable, if black holes are eternal, or transforming an arbitrary amount of baryons into a state
with zero baryonic number. For instance
a black hole consisting entirely from baryons would completely evaplorate creating (almost) equal number
of baryons and antibaryons. In the process of evaporation a small baryon asymmetry might be created 
but it normally would be negligibly small in comparison with the initial baryonic number captured at the black hole
formation.

If baryonic number is conserved in particle interations, the proton must be almost absolutely stable. 
To be more precise it may decay by Zeldovich mechanism~\cite{YaZ-decay}
 through formation of a virtual black hole from three quarks inside proton. But the life-time
 with respect to such decay is almost infinite, $\tau_p \sim 10^{45}$ years. Also one could hardly expect 
neutron-antineutron oscillations  induced by virtual BHs to be observable (for a recent review see e.g. \cite{n-anti-n}).

Another unusual feature of the model is a possibility to create baryon (or any other type of asymmetry between
particles and antiparticles) in thermal equilibrium. Normally the deviation from thermal equilibrium is 
suppressed by the factor of the order of the ratio
of the Hubble expansion rate to the particle reaction rate, $H/\Gamma$. The former is inversely proportional
to a huge value of the Planck mass, $H \sim T^2/m_{Pl}$, where 
$T$ is the cosmological plasma temperature. According to the estimates presented above,
for the mechanism considered here the situation is opposite: the larger is the Planck mass 
(or the slower is the 
cosmological expansion), the larger is the baryon asymmetry. On the other hand, the gravitational
attraction which forces massive $X$-particles to fall on the nearest BH is inversely proportional
to $m_{Pl}^2$, so ultimately the effect disappears in the limit of infinite $m_{Pl}$  as well.

The magnitude of the baryon asymmetry evidently strongly depends upon the cosmological expansion 
regime. In particular, it would be very interesting to study baryogenesis
in the frameworks of $R^2$ inflation~\cite{aas-infl} there exists a long period of
the universe evolution during which the fall off of the cosmological 
temperature is drastically different from that accepted in the conventional 
cosmology~\cite{EA-AD-RS-1,ADS-symmetry}. 

In the course of working on the presented here version of baryogenesis we became aware of an 
interesting modification of the scenario presented in ref.~\cite{DP-BH-BG} on
the generation of the cosmological baryon  asymmetry through the capture of antibaryons  by PBH~\cite{napoli}, 
which also may lead to an efficient baryogenesis.

\section*{Acknowledgment}

This work was supported by RSF Grant 20-42-09010. 

The Feynman diagrams was drawn by JaxoDraw~\cite{jaxo}.

\end{document}